\documentclass[]{aa}
\usepackage{psfig}
\begin{document}

\thesaurus{08(08.15.1)}

\title{V2109~Cygni, a second overtone field RR Lyrae star}

\author{L.L. Kiss\inst{1} \and B.Cs\'ak\inst{1} \and J.Thomson\inst{2}
\and J.Vink\'o\inst{3}}

\institute{Department of Experimental Physics and Astronomical Observatory,
JATE University,
Szeged, D\'om t\'er 9., H-6720 Hungary \and
David Dunlap Observatory, University of Toronto, Richmond Hill, Canada
\and
Department of Optics \& Quantum Electronics, JATE University, Research
Group on Laser Physics of the Hungarian Academy of Sciences}

\titlerunning{V2109 Cyg, a second overtone RR Lyrae}
\authorrunning{Kiss et al.}
\offprints{l.kiss@physx.u-szeged.hu}
\date{}

\maketitle
 
\begin{abstract}

We present the first (as of February, 1999) $UBV$ and $uvby$ photometric
measurements for the short period variable star V2109~Cyg discovered
by the Hipparcos satellite and classified as a field RRc variable.
We have obtained new times of maxima and the period change
has been studied. We determined the fundamental physical parameters
using the geometric distance of the star and the most recent
synthetic colour grids. The results are:

\noindent [Fe/H]=$-0.9\pm0.2$, $M_{bol}=0.73\pm0.43$ mag,
$\langle T_{\rm eff}\rangle=6800\pm200 K$, ${\rm log g}=2.7\pm0.2$,
$R=4.6\pm0.9 R_{\odot}$.

Additionally, we took medium resolution ($\lambda/\Delta
\lambda\approx 11000$) spectra in the red spectral region
centered at 6600 \AA. A complete radial
velocity curve has been determined from 60 spectra.
We found no systematic difference between the velocities
from H$\alpha$ and metallic lines indicating a smooth
pulsation. The mass of the star
is found to be $M=0.5\pm0.3 M_{\odot}$ using the photometric
$log g$ value and the acceleration curve calculated
from the radial velocity data. The finally adopted set of
the physical parameters lies in the typical range of RR~Lyrae
stars, which implies that V2109~Cyg is the shortest period
RR~Lyrae-type variable.      
The visual amplitude vs. period and the Fourier amplitude parameter
$R_{\rm 21}$
vs. period diagrams suggest that V2109~Cyg probably pulsates
in the second overtone mode.

 
\end{abstract}

\section{Introduction}

V2109~Cyg (=HD191635, $\langle V \rangle=7.49$, $\Delta V=0.16$,
$P=0\fd18605$)
was discovered by the Hipparcos satellite (ESA 1997) and
was classified as an RRc variable based on the observed properties.
Its period is shorter than that of
the shortest period RR~Lyrae variable
in the 4th edition of the General Catalogue of Variable Stars (V1407~Sgr,
RRc, $P=0.\!\!^{\rm d}218689$). Furthermore, its brightness places the star
among the brightest RR Lyrae-type variables known. In fact, only RR~Lyrae
itself is brighter around maximum light and S~Eri, which
classification is quite uncertain (S~Eri -- RRc:, $P=0\fd27$, spectral
type F0IV).

The short period of V2109~Cyg falls far beyond the typical range observed
in RRc variables.
Several authors claimed for similarly short-period (between 0$.\!\!^{\rm d}$21
-0$.\!\!^{\rm d}$28) RRc stars in globular clusters that they may be
second overtone RR Lyrae (RRe) variables (e.g. Clement et al. 1979,
Walker 1994, Walker \& Nemec 1996). Stothers (1987) concluded, based on
his hydrodinamical calculations, that second overtone pulsators
probably do not exist among RR Lyrae type stars.
Recently, Kov\'acs (1998) argued
the second overtone interpretation of the observed short-period
RRc stars in globular clusters and in LMC (Alcock et al. 1996)
using the Fourier decomposition of the suspected
RRe stars. He suggested that the earlier candidate RRe stars are
ordinary RRc variables at the short-period end of the
instability strip.

We started a long-term observational project of Str\"omgren photometry
of the newly discovered bright Hipparcos variables in 1998.
We have chosen V2109~Cyg because of its brightness, period and
theoretical importance. Since
the period, spectral type and light curve amplitude alone do not
exclude the possibility of wrong classification (i.e. there are
a few high-amplitude $\delta$ Scuti stars with similar photometric
parameters -- see e.g. Kiss et al. 1999, Rodr\'\i guez et al. 1998),
accurate determination of the fundamental physical
parameters is highly desirable.

The main aim of this paper is to present the first $UBV$ and $uvby$
photometry for V2109~Cyg. Also, our radial velocity measurements are
the first time-resolved spectroscopic observations of this star to date.
The paper is organised as follows: the photometric and spectroscopic
observations are described in Sect. 2. Sect. 3 deals with the detailed
multicolour photometric analysis and the determination of fundamental
physical parameters, while the mode of pulsation is discussed in Sect. 4.

\section{Observations}

\begin{table}
\begin{center}
\caption{The journal of observations}
\begin{tabular} {ll}
\hline
Julian Date & type\\
\hline
2451032    & $uvby$\\
2451037    & $uvby$\\
2451080    & $uvby$\\
2451087    & $UBV$\\
2451106    & spectr.\\
2451108    & spectr.\\
2451110    & $UBV$\\
\hline
\end{tabular}
\end{center}
\end{table}

\subsection{Photometry}

The photometric observations were carried out on 5 nights during
August-November, 1998 using the 0,4 m Cassegrain-type telescope of
Szeged Observatory. The detector was a single-channel Optec
SSP-5A photoelectric
photometer equipped with $UBV$ and $uvby$ filters supplied by the
manufacturer.
We made differential photometry in respect to HD191022 ($V$=7.44, $B-V$=0.66,
$b-y$=0.41, $m_{\rm 1}$=0.20, $c_{\rm 1}$=0.38). The resulting overall
accuracy is about $\pm0.01$ for $V$ and $B-V$, $\pm0.015$ for
$U-B$ and $b-y$, $\pm0.025$ for $m_{\rm 1}$ and $c_{\rm 1}$. This
considerably large scatter is due to the frequent atmospheric
instabilities caused by the nearby city of Szeged. Fortunately the
large number of individual points (468 in $uvby$, 178 in $UBV$ colours)
allowed the calculation of accurate normal curves\footnote{Individual
photometric data are available upon request from the first author
(l.kiss@physx.u-szeged.hu).}.
The light, colour and radial velocity curves (Sect.2.2.) were phased using
the finally adopted period $P=0\fd186049$ and epoch
Hel.JD(max.)=2451032.3936 and can be seen in Fig.1. (see Sect. 3.1. for
the period analysis).

\begin{figure}
\begin{center}
\leavevmode
\psfig{figure=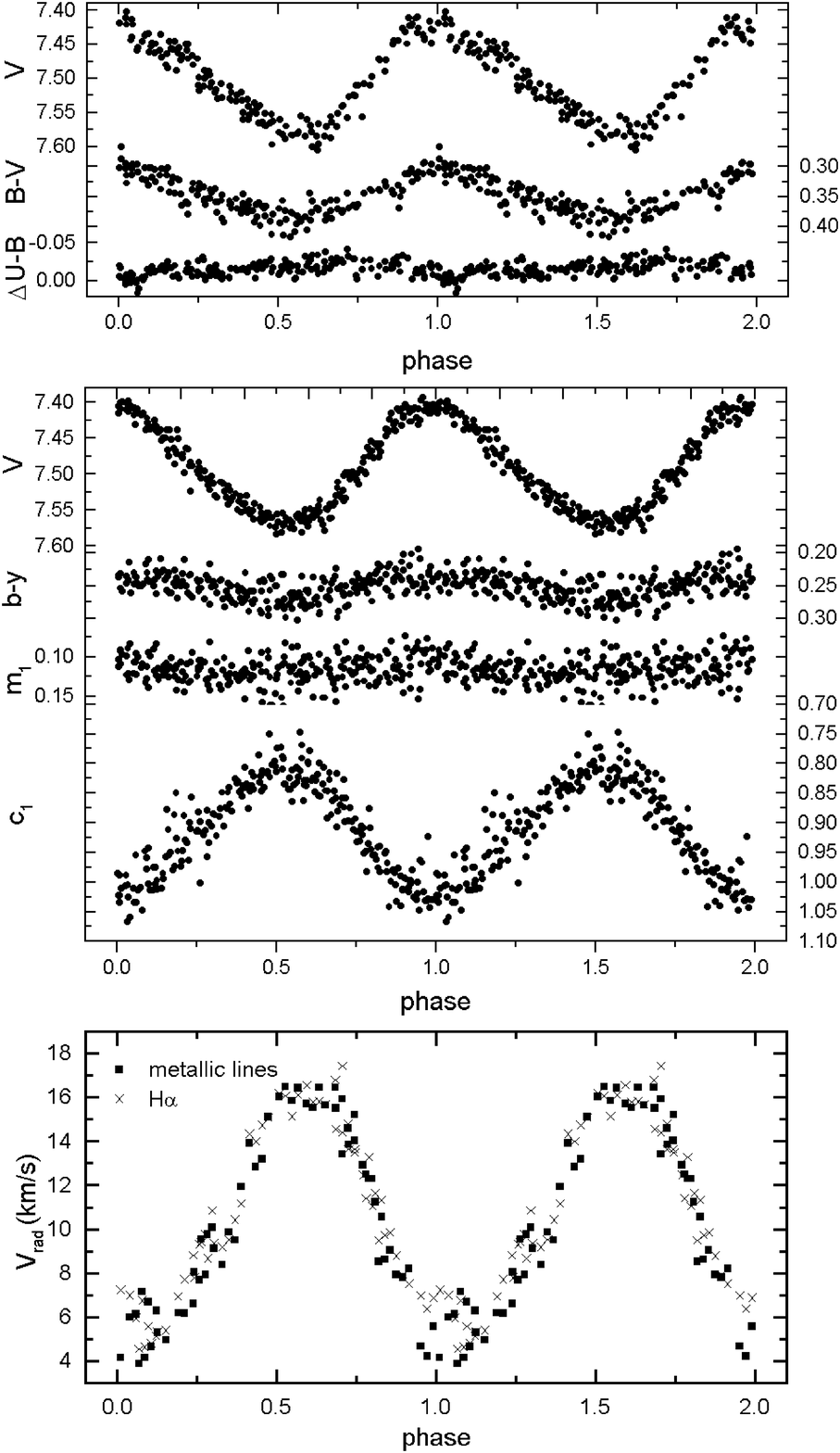,width=\linewidth}
\caption{The light, colour and radial velocity curves phased according
to the adopted ephemeris.}
\end{center}
\label{allph}
\end{figure}

\subsection{Spectroscopy}

The spectroscopic observations were carried out at David Dunlap Observatory
with the Cassegrain spectrograph attached to the 74" telescope. The
detector and the spectrograph setup was the same as used by 
Vink\'o {\it et al.} (1998). The resolving power ($\lambda/\Delta \lambda$)
was 11,000 and the signal-to-noise ratio reached about 60.

\begin{table*}
\begin{center}
\caption{The observed heliocentric radial velocities from the two 
cross-correlated spectral region.}
\begin{tabular} {lrrlrrlrr}
\hline
Hel. J.D.  &  $V_{\rm rad(H\alpha)}$ & $V_{\rm rad(m.)}$ & Hel. J.D.  &  $V_{\rm rad(H\alpha)}$ & $V_{\rm rad(m.)}$ & Hel. J.D.  &  $V_{\rm rad(H\alpha)}$ & $V_{\rm rad(m.)}$ \\
\hline
2451106.4859 &  7.8   &  8.1  &  2451106.5687 &  14.6  &  15.5  &    2451106.8044  & 7.0 &  4.7\\
2451106.4902 &  9.4   &  9.6  &  2451106.5723 &  14.4  &  13.4  &    2451106.8081  & 6.4 &  4.2\\
2451106.4938 &  8.7   &  9.8  &  2451106.5760 &  13.7  &  13.8  &    2451106.8117  & 6.9 &  5.6\\
2451106.4975 &  9.4   &  9.1  &  2451106.5796 &  13.7  &  14.0  &    2451106.8153  & 7.2 &  4.1\\
2451106.5025 &  9.2   &  8.4  &  2451106.5863 &  11.4  &  12.5  &    2451106.8203  & 7.0 &  6.0\\
2451106.5062 &  9.5   &  9.9  &  2451106.5899 &  11.1  &  12.3  &    2451106.8240  & 6.0 &  6.1\\
2451106.5098 &  10.4  &  9.5  &  2451106.5936 &  9.5   &  8.5   &    2451106.8276  & 6.7 &  7.1\\
2451106.5135 &  11.2  &  11.9 &  2451106.5972 &  9.7   &  8.6   &    2451106.8313  & 5.6 &  6.7\\
2451106.5183 &  14.3  &  13.9 &  2451106.7543 &  16.8  &  16.4  &    2451106.8360  & 5.1 &  6.3\\
2451106.5220 &  14.0  &  12.8 &  2451106.7584 &  17.4  &  15.9  &    2451108.5002  & 4.5 &  3.9\\
2451106.5256 &  14.7  &  13.1 &  2451106.7620  & 14.8 & 14.5    &    2451108.5038  & 4.6 &  4.2\\
2451106.5293 &  15.1  &  15.1 &  2451106.7657  & 13.5 & 15.2    &    2451108.5074  & 4.8 &  4.7\\
2451106.5356 &  16.2  &  16.0 &  2451106.7705  & 12.5 & 12.9    &    2451108.5111  & 5.3 &  5.3\\
2451106.5393 &  16.1  &  16.4 &  2451106.7741  & 13.3 & 12.2    &    2451108.5160  & 5.4 &  5.0\\
2451106.5429 &  15.1  &  15.8 &  2451106.7778  & 11.7 & 11.2    &    2451108.5233  & 6.9 &  6.2\\
2451106.5466 &  16.1  &  16.4 &  2451106.7814  & 11.3 & 10.5    &    2451108.5269  & 7.7 &  6.2\\
2451106.5516 &  16.6  &  15.7 &  2451106.7864  & 9.8  & 9.0     &   2451108.5320  & 8.8  & 6.6 \\
2451106.5552 &  15.8  &  15.5 &  2451106.7901  & 8.8 &  7.9     &   2451108.5356  & 9.3  & 7.7 \\
2451106.5589 &  15.8  &  16.4 &  2451106.7937  & 7.9 &  7.8     &    2451108.5393  & 9.8 &  7.9\\
2451106.5625 &  15.7  &  15.6 &  2451106.7974  & 7.5 &  8.2     &    2451108.5429  & 10.8 &  10.0\\
\hline
\end{tabular}
\end{center}
\end{table*}

The spectra were reduced with standard IRAF tasks, including bias
removal, flat-fielding, cosmic ray elimination, order extraction (with
the task {\it doslit}) and wavelength calibration. For the latter, two FeAr
spectral lamp exposures were used, which were obtained immediately
before and after every four stellar exposures. The sequence
of observations FeAr-var-var-var-var-FeAr was chosen because of the
short period of V2109~Cyg. Careful linear interpolation between the
two comparison spectra has been applied in order to take into
account the sub-pixel shifts of the four stellar spectra caused by
the movement of the telescope. The exposure time was fixed as 5 minutes,
which corresponds to 0.02 in phase, avoiding phase smearing of
the radial velocity curve.
The spectra were normalized to the continuum by fitting cubic spline,
omitting the region of H$\alpha$.

Radial velocities were determined by cross-correlating parts of the
spectra of V2109~Cyg with the spectrum of the IAU standard
velocity star HD187691, using the IRAF task {\it fxcor}. The spectral
type and radial velocity of HD187691 are F8V and $+0.1\pm0.3$ km s$^{-1}$.
The cross-correlated regions were [6550--6580] (H$\alpha$)
and [6580--6700] \AA\ (photospheric metal lines). Typical spectra
of V2109~Cyg and HD187691 are presented in Fig.2.

\begin{figure}
\begin{center}
\leavevmode
\psfig{figure=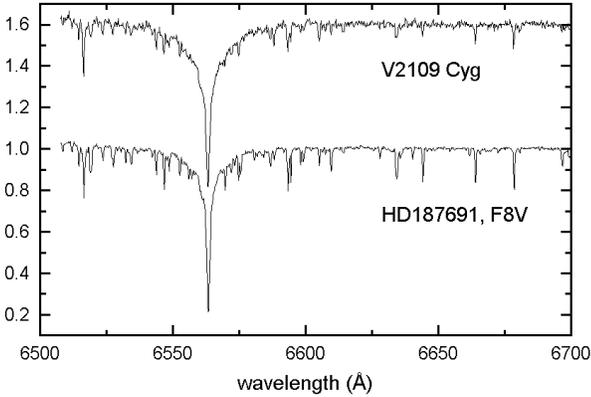,width=\linewidth}
\caption{Sample spectra of V2109~Cyg and HD187691.}
\end{center}
\label{spectra}
\end{figure}

The observed heliocentric radial velocities are presented in Table 2.
The velocimetric accuracy is about $\pm1-1.5$ km s$^{-1}$, which
was estimated by the residual scatter of the measurements around a
fitted low-order Fourier polynomial. This equals 0.1 pixel uncertainty
which could be associated with the limited accuracy of the wavelength
calibration, as well as the slightly different spectral types 
of V2109~Cyg and HD187691. Fortunately, the latter affects the resulting
radial velocities less significantly, as almost all of the metallic lines
in the spectrum of HD187691 can also be identified in V2109~Cyg.

We have found no significant systematic difference between the 
velocities of the two
cross-correlated regions, which means that the velocity gradient
between the atmospheric regions of H$\alpha$ and metallic lines
does not exceed our
accuracy (see Vink\'o et al. 1998 concerning this comparison for
Type II Cepheid variables).

\section{Photometric analysis}

\subsection{Period}

The photometric period was studied by means of standard Fourier analysis
and the classical O$-$C diagram. We have calculated the Discrete Fourier
Transform for all measurements made through V filter.
The DFT spectrum showed a principal peak at $f_{\rm 0}=5.37488$ c/d
(P=0\fd18605), in good agreement with the Hipparcos-period
(0\fd1860656). After prewhitening with this frequency
and its harmonics, the resulting periodogram did not contain any
significant peak (Fig.3). This suggest the monoperiodic nature of V2109~Cyg.

\begin{table}
\begin{center}
\caption{New times of maxima of V2109~Cyg.}
\begin{tabular} {ll}
\hline
i &  T$_{i}$(HJD)\\
\hline
1 &  2451032.3936\\
2 &  2451037.4134\\
3 &  2451080.3928\\
4 &  2451110.3465\\
\hline
\end{tabular}
\end{center}
\end{table}

\begin{figure}
\begin{center}
\leavevmode
\psfig{figure=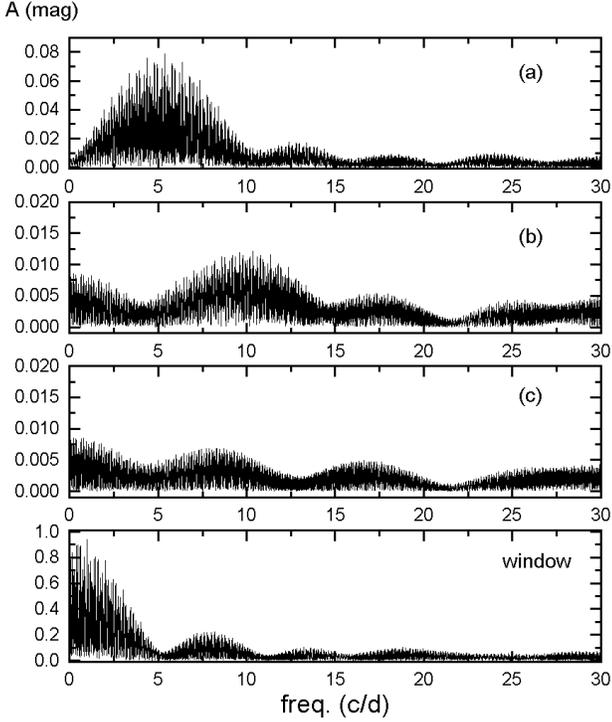,width=\linewidth}
\caption{The DFT amplitude spectrum with a principal peak $f_{\rm 0}$ (a),
the prewhitened spectrum with $f_{\rm 1}=2f_{\rm 0}$ (b) and the
remaining periodogram with no significant peak (c). The window function
is shown separately (bottom panel).}
\end{center}
\label{v2109wh}
\end{figure}

We have obtained four times of maxima using $V$ light curves (Table 3).
An O$-$C analysis was performed in order to refine the period value
(see a recent application of this technique in Kasz\'as {\it et al.} 1998).
The original
ephemeris from Hipparcos photometry (E$_{0}$=2448500.0280,
P=0\fd1860656) was used
to construct the top panel in Fig.4 (the error bars associated
to our data points correspond to a $\pm$0\fd002 uncertainty,
this relatively high value is due to the flat top of the light curve).
The positive shift in this diagram suggest a somewhat {\it longer} period
than that published by ESA (1997). The middle panel shows the
recalculated O$-$C values using a period corrected with the slope of the
top diagram (P=0\fd1860662(3)).
However, there can be seen some hints of systematic decreasing trend
in our new O$-$C points, suggesting a {\it shorter} period. The corrected
O$-$C diagram is presented in the bottom panel of Fig.4
(P=0\fd186049(5)). The last period
is in better agreement with the result of the Fourier analysis, therefore
we conclude that the recent period is
0\fd186049$\pm$0\fd000005, and, assuming that the
Hipparcos data were phased properly, a sudden period decrease happened 
between 1991 and 1998. Follow-up observations are needed to
clarify the present state of period change.

\begin{figure}
\begin{center}
\leavevmode
\psfig{figure=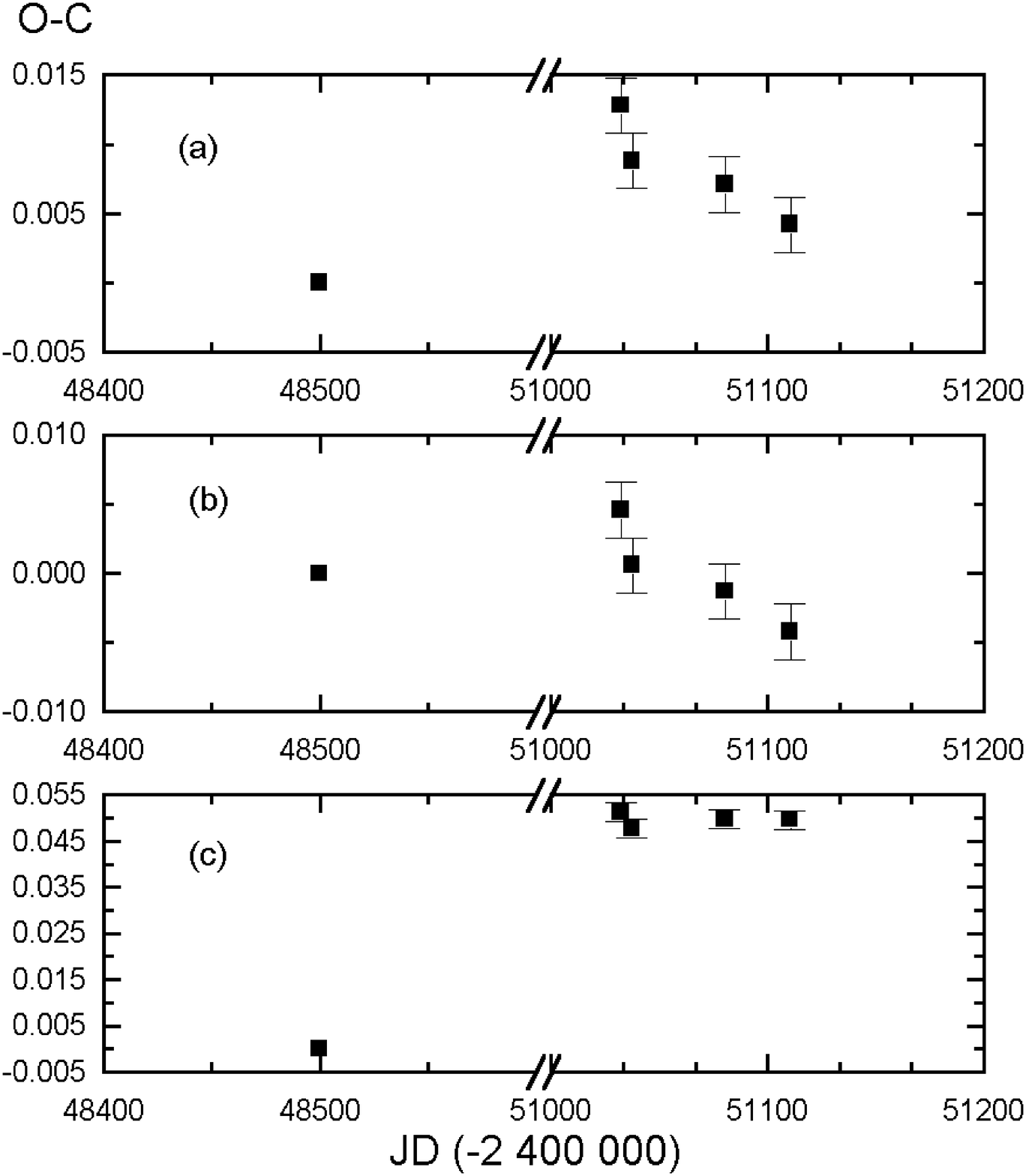,width=\linewidth}
\caption{The O$-$C diagram of V2109~Cyg with three ephemerides. (a): original
ephemeris; (b): assuming constant period over the time base;
(c): assuming sudden period jump between the older and recent observations
(note the break in the horizontal axis and large positive O$-$C values).}
\end{center}
\label{v2109oc}
\end{figure}

\subsection{Str\"omgren calibrations -- physical parameters}

In order to minimize the effect of the observational scatter, we used
the normal curves (Table 4) of the light and colour variations to
infer physical parameters of V2109~Cyg.

\begin{table}
\begin{center}
\caption{Str\"omgren photometry (normal points) of V2109~Cyg.}
\begin{tabular} {lcccc}
\hline
phase & $V$ & $b-y$ & $m_{\rm 1}$ & $c_{\rm 1}$\\
\hline
0.00 & 7.407 & 0.242 & 0.104 & 1.017\\
0.05 & 7.414 & 0.240 & 0.109 & 1.007\\
0.10 & 7.430 & 0.237 & 0.113 & 0.998\\
0.15 & 7.450 & 0.242 & 0.115 & 0.962\\
0.20 & 7.468 & 0.246 & 0.119 & 0.927\\
0.25 & 7.493 & 0.247 & 0.118 & 0.919\\
0.30 & 7.512 & 0.252 & 0.120 & 0.901\\
0.35 & 7.530 & 0.260 & 0.124 & 0.864\\
0.40 & 7.544 & 0.265 & 0.122 & 0.838\\
0.45 & 7.557 & 0.263 & 0.125 & 0.821\\
0.50 & 7.565 & 0.265 & 0.123 & 0.814\\
0.55 & 7.565 & 0.270 & 0.118 & 0.815\\
0.60 & 7.560 & 0.270 & 0.119 & 0.819\\
0.65 & 7.549 & 0.264 & 0.119 & 0.840\\
0.70 & 7.529 & 0.260 & 0.116 & 0.867\\
0.75 & 7.497 & 0.254 & 0.112 & 0.906\\
0.80 & 7.473 & 0.243 & 0.114 & 0.941\\
0.85 & 7.439 & 0.236 & 0.117 & 0.983\\
0.90 & 7.419 & 0.242 & 0.111 & 1.007\\
0.95 & 7.411 & 0.245 & 0.104 & 1.008\\
\hline
\end{tabular}
\end{center}
\end{table}

The geometric distance of V2109~Cyg, as measured by the Hipparcos
satellite, is 205$\pm$40 pc. To convert the distance and the
apparent magnitude to absolute magnitude, one has to include the
effect of interstellar reddening.

The colour excess was estimated: {\it i)}
using a spectral type-colour relation for an
F0 star; {\it ii)} using intrinsic colour calibrations based on the
Str\"omgren indices.

The dereddened $B-V$ colour of an F0-type star is $(B-V)_{\rm 0}$=0.30
(Carroll \& Ostlie 1996). The mean $B-V$ colour of V2109~Cyg is 0.35$\pm$0.01
mag, thus, a resulting colour excess is $E(B-V)$=0.05. This value
has quite large uncertainty (at least $\pm$0.03 mag), since the spectral
type cannot be determined as precise as $\pm$2 subclass and consequently,
the $(B-V)_{\rm 0}$ might differ $\pm$0.03-0.05 mag.

Two Str\"omgren calibrations were applied to obtain the intrinsic $(b-y)_{\rm 0}$
colour. Crawford (1975) and Olsen (1988) presented $uvby-\beta$ calibrations
for F-type stars. Since we have no $\beta$-measurements,
we have to estimate the actual value of $\beta$ index using the different
relations of Str\"omgren indices vs. $\beta$ (see Figs.1, 3 and 6 in Crawford
1975). All of them suggest $\beta\approx2.70$ with $\pm$0.02 uncertainty.
Adopting this value, final calibration equations of Crawford (1975) and
Olsen (1988) give $E(B-V)$=0.00 and $E(B-V)=-$0.02, respectively
(a ratio of $E(b-y)/E(B-V)\approx$0.7 was used).
The simple average of these reddenings is $E(b-y)$=0.006, while a
weighted mean assigning a larger weight to the spectral-type--colour
relation (3-1-1 - because of the very uncertain determination
of the $\beta$ index) is $E(B-V)$=0.03$\pm$0.02.
Fortunately, the star lies
in a considerably close vicinity and, consequently, its reddening
is not expected to have a large value. The total visual absorption, $A_{\rm V}$,
due to this reddening probably does not exceed 0.10 mag.

The calculated
visual absolute magnitude corrected for interstellar absorption
is 0.83$\pm$0.43 mag, while
the bolometric absolute magnitude (BC(F0)=$-$0.10, Carroll \& Ostlie 1996)
is 0.73$\pm$0.43
(the uncertainty is mainly due to the error of the distance). This
gives 41$\pm$15 $L_{\odot}$ for the luminosity of the star.

The metallicity, expressed with the [Fe/H] value, was determined by
Eq. (2) of Malyuto (1994), which involves $b-y$ and $m_{\rm 1}$ indices.
The mean $b-y$ (=0.255) and $m_{\rm 1}$ (=0.112)
values give [Fe/H]=$-$0.9$\pm$0.2 for V2109~Cyg,
which suggests a relatively metal rich field RR~Lyrae star. The high
luminosity also strengthens the classification (see below for further
discussion).

The atmospheric parameters $T_{\rm eff}$ and $log~g$ were obtained
using the most recent synthetic colour grids of Kurucz (1993). The
use of $(b-y)_{\rm 0}$--($c_{\rm 1})_{\rm 0}$ colour-colour diagram
is shown
in Fig.5, where the normal points presented in Table 4 are plotted.
The corresponding T$_{\rm eff}$--log~g pairs were determined by a
two-dimensional linear interpolation, resulting in
$\langle T_{\rm eff} \rangle = 6800\pm200$ K and
$\langle log~g \rangle = 2.7\pm0.2$ dex. It can be seen in Fig.5 that
the points run almost parallel with the lines of constant temperature,
indicating that the colour variation is mainly caused by the
changing gravity, not the variation of the temperature.

\begin{figure}
\begin{center}
\leavevmode
\psfig{figure=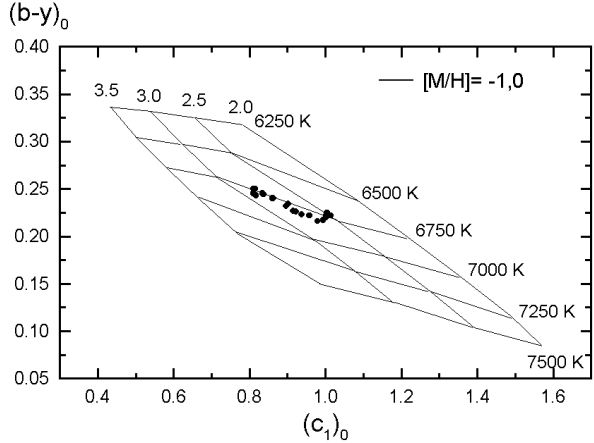,width=\linewidth}
\caption{The observed colour-colour variations plotted on the
synthetic grid of Kurucz (1993).}
\end{center}
\label{byc10}
\end{figure}

Combining the mean temperature, luminosity and solar values
$T_{\rm eff}=5770$ K and $M_{\rm bol}$=4.75 (Allen 1976),
we calculated the
mean stellar radius $R_*=4.7\pm0.9$ R$_{\odot}$. By means of
effective gravity

\begin{center}
$$g_{\rm eff}=G {M_* \over R_*^2}+p {dV_{\rm r} \over dt}$$
\end{center}

\noindent where p=1.36 is the geometric projection factor
(e.g. Burki \& Meylan 1986), we obtain a stellar
mass of $M_*=0.5\pm0.3$ M$_{\odot}$. In summary, therefore, we adopt

\bigskip

\noindent $M_V=0.83\pm0.43$ mag

\noindent $M_{\rm bol}=0.73\pm0.43$ mag

\noindent $L=41\pm15 L_{\odot}$

\noindent [Fe/H]=$-0.9\pm0.2$

\noindent $\langle T_{eff} \rangle = 6800\pm200$ K

\noindent $\langle {\rm log~g} \rangle = 2.7\pm0.2$ dex

\noindent $\langle R_* \rangle = 4.6\pm0.9 R_{\odot}$

\noindent $M_*=0.5\pm0.3 M_{\odot}$

\section{V2109~Cyg: a second overtone RR~Lyrae}

V2109~Cyg was classified as an RRc star based on the Hipparcos data.
However, both the period and the shape of the V light curve resemble
to those of a high-ampitude $\delta$ Scuti star, too. But the finally
adopted set of the physical parameters lies in the typical
range of RR~Lyrae properties. For instance, one of the brightest
and nearest RRc stars, DH~Peg (Fernley et al. 1990)
has almost the same parameters than V2109~Cyg (DH~Peg:
$\langle T_{eff} \rangle = 7070\pm210$ K, $\langle R_* \rangle = 4.4\pm0.9 R_{\odot}$,
$M_{bol}=0.65\pm0.40$).

\begin{figure}
\begin{center}
\leavevmode
\psfig{figure=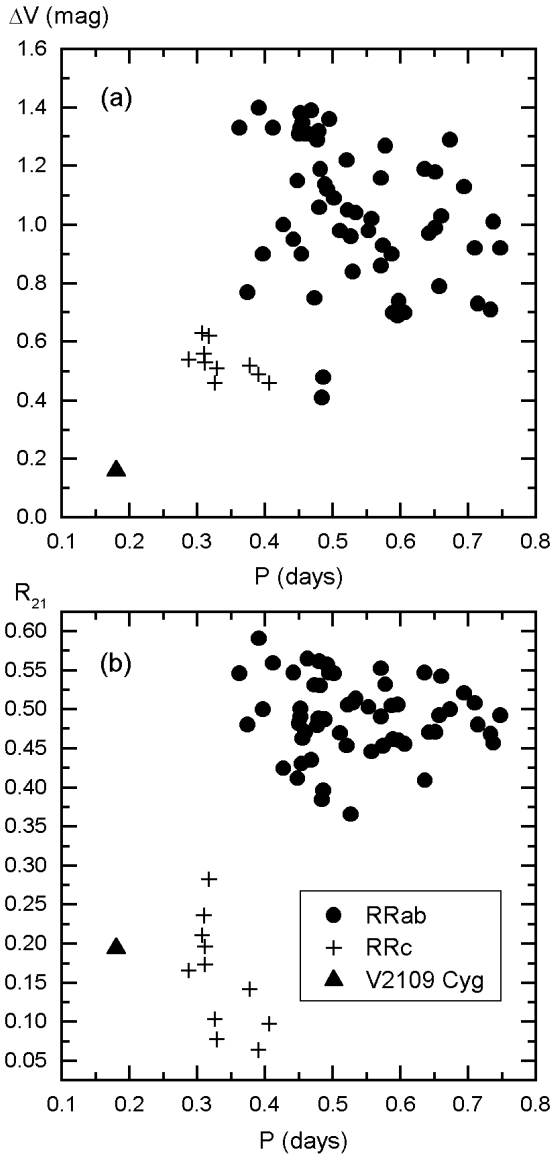,width=\linewidth}
\caption{A comparison of the visual amplitude vs. period (a) and
Fourier amplitude parameter R$_{\rm 21}$ vs. period (b) for V2109~Cyg
and field RRab and RRc variables.}
\end{center}
\label{v2109fou}
\end{figure}

The relatively large radius and low mass also exclude the dwarf Cepheid
possibility, since typical values of $\langle R_* \rangle$ and $M_*$ for
these stars are 1.5--3.0 $R_{\odot}$ and 1--1.5
$M_{\odot}$ (Fernley {\it et al.} 1987). Also the small
colour variation indicates
very small change in the effective temperature during the pulsational
cycle. Both high-amplitude $\delta$~Scuti and SX~Phoenicis stars
exhibit at least 600-800~K variation, while the difference
between the maximum and minimum photometric temperature in V2109~Cyg is only
100~K.

As has been mentioned in the Introduction, the period of V2109~Cyg
is shorter than the shortest period of an RR~Lyrae star listed in
GCVS. This raises the question of the mode of the pulsation, since there
are some pieces of evidence for ultra-short period RR~Lyrae stars
identified as candidate RRe stars in some globular clusters (e.g Walker \&
Nemec 1996).

We plot the the visual amplitude of V2109~Cyg versus the period
in Fig.6a with field RRab and RRc stars taken from Simon \& Teays
(1982). Accepting that RRab and RRc stars are fundamental
and first-overtone pulsators, the position of V2109~Cyg strongly
suggests a possibility of a second overtone variation. Fig.6b
shows the Fourier amplitude ratio parameter R$_{\rm 21}$
(0.19 for V2109~Cyg) for the same sample.
Again, V2109~Cyg lies far from the two populated
regions of RRab and RRc variables.
The mentioned value for R$_{\rm 21}$ was calculated from the Johnson V data,
as Simon \& Teays (1982) used V-observations. The Str\"omgren y (transformed
to V) data give slightly different result of R$_{\rm 21}$=0.12, but the
difference does not affect our conclusions.
 Fig.6a has essentially
the same structure than e.g. Figs.2-3 of Sandage (1981), where
this distribution is plotted for two globular clusters (M3 and
$\omega$ Cen), illustrating that
it is a general property of RR~Lyrae variables.
To our knowledge, there is no similar study concerning the Fourier
decomposition of the observed radial velocity curves, only
Simon (1985) studied theoretical velocity curves of different
pulsational modes. Plotting R$_{\rm 21}$(rad. vel.)=0.13 for
V2109~Cyg in Fig.5 of Simon (1985), the position is similarly
deviant from the first-overtone models as in the photometric case.

Furthermore, we can compute the pulsational constant, although its
usefulness is questionable being a complex function of mass, temperature,
luminosity, metallicity, etc. For $P=0.18605$, $R=4.7\pm0.9 R_{\odot}$ and
$M=0.5\pm0.3 M_{\odot}$ we get $Q=P(M/R^3)^{1/2}=0.013\pm0.009$.
This small pulsational constant could be a sign 
of the second overtone mode (we recall Fernley et al. 1990
for a similar discussion on the real nature of DH~Peg).

Finally, the second overtone interpretation fits very well the recent
theoretical results by Kov\'acs (1997) and Kov\'acs (1998). Kov\'acs (1997)
examined the possibility of the existence of second overtone RR~Lyrae stars
employing the results of linear non-adiabatic (LNA) and evolutionary
calculations. He failed to explain the 0.28 day peak in the MACHO
RR~Lyrae sample (Alcock et al. 1996) as a result of the second overtone
stars, since the different parameter combinations and the calculated
growth rates of the first and second overtone modes excluded
strongly excited
second overtone mode with period around 0.28 day. However, LNA
calculations suggested strong excitation of the second overtone mode
for short ($\leq$0.23 days) period with high temperature and low luminosity.
On the other hand, combining the evolutionary parameters with
the results of LNA calculations, Kov\'acs (1997) concluded that
the only possible scenario is the one when the star evolves from
low temperatures and looses its stability in the fundamental
pulsation. One can find different parameter settings where
both the first and second overtones have stable limit cycles and
it is possible that the second overtone switches on instead of the
first one. In the second paper, Kov\'acs (1998) tested the second overtone
candidates in M68 and IC~4499 and found that in a parameter regime
of $M/M_{\odot}=0.65\pm0.20$, $L/L_{\odot}=50\pm20$ and
$T_{\rm eff}=6500\pm500$ the resulting second overtone periods
lie close to 0.22 days with an error of 0.03 days. Both period
and determined physical parameters of V2109~Cyg fall in the
ranges suggested by these theoretical calculations, strongly
supporting our hypothesis.

\section{Summary}

Based on the conclusions presented in the previous sections, we can
summarize our results:

1. We present the first continuous $UBV$ and $uvby$ photometry of the newly
discovered bright Hipparcos variable V2109~Cyg. We obtained
the radial velocity curve based on medium-resolution spectroscopy.
Our spectroscopic
observations clearly show that V2109~Cyg is indeed a
pulsating variable.

2. The reddening of the star was determined with different
photometric methods, resulting in E(B$-$V)=0.03$\pm$0.02.
The metallicity value from the Str\"omgren calibration of
Malyuto (1994) is [Fe/H]=$-0.9\pm0.2$, suggesting a relatively
metal rich field RR~Lyrae star.

3. Using the parallax measured by the Hipparcos satellite and the
most recent synthetic grids of Kurucz (1993), we have determined
the fundamental physical parameters (luminosity, effective
temperature and surface gravity, mean radius, mass) which are
typical among the RR~Lyrae variables.

4. We suggest that V2109~Cyg probably pulsates in the second
overtone. This conclusion is based on the 
following pieces of evidence. First, its period 
is the shortest one among the
classified RR~Lyrae stars. Second, the occupied position on the
$\Delta V$ vs. period and R$_{\rm 21}$ vs. period diagrams
is highly deviant both from the RRc and RRab variables.
This deviation is in the same direction and distance where one
can expect the location of the second overtone pulsators.
Third, recent theoretical results concerning the second
overtone pulsation in RR~Lyrae stars suggested such a parameter
range that corresponds exactly to those ones derived for
V2109~Cyg.
The small value of the pulsational constant (Q=0.013) also
supports this hypothesis.

\begin{acknowledgements}
This research was supported by Hungarian OTKA Grants \#F022249,
\#T022259, Grant PFP~5191/1997, Szeged Observatory Foundation
and Foundation for Hungarian Education and Science (AMFK).
Fruitful discussions with K. Szatm\'ary are gratefully
acknowledged. The NASA ADS Abstract
Service was used to access data and references.
\end{acknowledgements}

\end{document}